\documentclass[aps,prl,showpacs,reprint,superscriptaddress,floatfix]{revtex4-2}
\usepackage{epsf,graphicx,color}
\begin{document}

\title{Extending the scope and understanding of all-optical magnetization switching in Gd-based alloys by controlling the underlying temperature transients} 

\author{Maxime Verges, Wei Zhang,  Quentin Remy, Yann Le-Guen, Jon Gorchon, Gregory Malinowski, Stephane Mangin, Michel Hehn, Julius Hohlfeld}
\affiliation{Institut Jean Lamour, CNRS UMR 7198, Université de Lorraine, Vandoeuvre-lès-Nancy F-54506,
France}

\date{\today}
\begin{abstract}

We use the thickness of Cu layers to control all-optical switching of magnetization in adjacent Gd$_{24}$(Fe$_{90}$Co$_{10})_{76}$ films. While increasing the Cu thickness from $5$ to $900\,$nm has no effect on the switching threshold, it significantly enlarges the fluence and pulse duration at which multiple domains emerge. Having shown that thermally activated multi-domain formation limits the maximum fluence and pulse duration for controlled switching, we demonstrate that continuous magnetization reversal precedes multi-domain formation in Gd$_{18}$Dy$_{4}$Co$_{78}$ films excited with fluences slightly larger than the multi-domain threshold.
\end{abstract}

\maketitle

Optical excitation of ferrimagnetic rare earth/transition metal (RE/TM) alloys and multi-layers by a single, arbitrary polarized laser pulse can lead to ultrafast magnetization reversal when the fluence and duration of this pulse stay within mutually dependent bounds \cite{GWY_16,DJM_20, WZH_21,GRW_13,LPH_17,RVS_11,SAH_11,JOL_21}. In Gd-based samples, this so-called  all-optical helicity independent switching (AO-HIS) (also known as temperature induced magnetization switching (TIMS)) results from transient spin transfer between the two magnetic subsystems by angular momentum conserving exchange scattering that originates from their different de- and remagnetization rates in response to laser pulse induced changes of the electron temperature. This understanding is the essence of so many experimental and theoretical investigations that it is impossible to do justice to each single one of them. Four experiments, showing the appearance of a transient ferromagnetic state during the reversal \cite{RVS_11} and magnetization reversal due to picosecond heat pulses resulting from current pulse induced joule heating \cite{YWG_17} and from fast heat diffusion through thick gold films \cite{XDM_17,WGY_17}, stand out as key demonstrations of the exchange and temperature driven nature of the magnetization reversal in GdFeCo, respectively.

Theoretical works have identified the dominance of spin transfer between the magnetic subsystems over spin transfer to the lattice as prerequisite for AO-HIS \cite{AOB_13,BLD_19,BBI_13,BAO_13,BaS_15,GOH_17,Gri_18,KaK_16,MHA_12,ScK_13,WHC_13}, derived preconditions for the magnetization magnitudes of both subsystems from this switching criterion \cite{JaA_22}, and qualitatively reproduced the impact of Gd-concentration, laser fluence and initial temperature on AO-HIS \cite{ABC_14,BAO_13}. However, the questions if the fluence thresholds for the appearance of controlled AO-HIS and of stochastic multi-domain formation are linked to thresholds for the maximum electron temperature \cite{RGA_22} or not \cite{GWY_16,JOL_21} and if the maximum pulse duration leading to AO-HIS is limited by a breakdown of the underlying switching mechanism \cite{DJM_20}, or by thermal erasure at longer time delays \cite{WZH_21} are still matter of debate.

\begin{figure}[hb] 
\includegraphics[width=8.5cm]{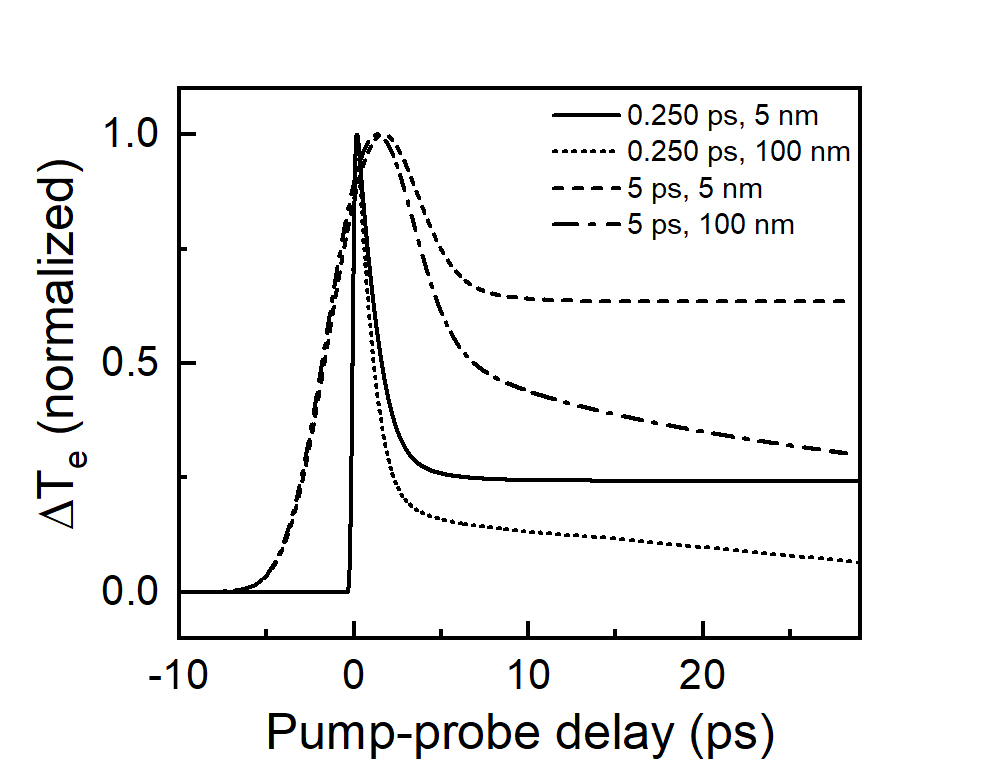}
\caption{\label{fig_HSprinciple} Transients of $\Delta T_{\rm e}/\Delta T_{\rm e,max}$ within GdFeCo predicted by the TTM for   $\tau_{\rm las}$ and $t_{\rm Cu}$
given in the legend.}
\end{figure}

In this letter, we use the thickness of Cu heat sinks, $t_{\rm Cu}$, to  control the transient  laser pulse induced increase of the electron temperature within adjacent $10\,$nm thick GdFeCo films, $\Delta T_{\rm e}(t)$, that results from their excitation by short laser pulses of various fluence, $F_{\rm las}$, and pulse duration, $\tau_{\rm las}$. This allows us to use slight adjustments of the laser fluence as means to tailor  $\Delta T_{\rm e}(t)$  for any given laser pulse duration such, that identical heating dynamics are followed by faster and stronger cooling for thicker heat sinks (cf.\,Fig.\ref{fig_HSprinciple}). 
In this way we are able to disentangle the impact of $\Delta T_{\rm e}$ at short times, where it reaches its maximum, and at longer times, where it is in quasi-equilibrium with the elevated phonon temperature, on the fluence thresholds for controlled switching, $F_{\rm SW}$, and of  stochastic multi-domain formation,  $F_{\rm MD}$. We derive both thresholds from domain images acquired via static linear magneto-optical microscopy.

The  investigated samples are grown by dc magnetron sputtering onto glass substrates in the following multilayered structure: glass/Ta(3)/Cu(5)/Gd$_{24}$(Fe$_{90}$Co$_{10})$(10)/Cu($t_{\rm Cu}$)/Pt(5), where the numbers in brackets give the layer thickness in nm. The bottom Ta- and Cu-layers improve adhesion of the structure to the glass substrate and ensure that the GdFeCo-layer exhibits perpendicular magnetic anisotropy, respectively \cite{BLM_22}. They are thin enough to allow probing of the magnetic state of the GdFeCo layer via measurements of the linear magneto-optical Kerr effect from the substrate side. The Pt-capping layer prevents oxidation of the sample. The thickness of the upper Cu heat sink wedges,  $t_{\rm Cu}$,  varies linearly across the 2.5\,cm wide samples over the following ranges (in nm): [5,10], [10,15], [10,20], [20,30], [35,70], [70,105], [50,100], [100,150], [100,200], [200,300], [300,600], and [600,900]. Great care was taken to ensure identical compositions/magnetic properties of all GdFeCo layers, which was confirmed by the demonstration of identical hysteresis loops and equal fluence thresholds for different samples with equal $t_{\rm Cu}$.  

\begin{figure}[ht] 
\includegraphics[width=8cm]{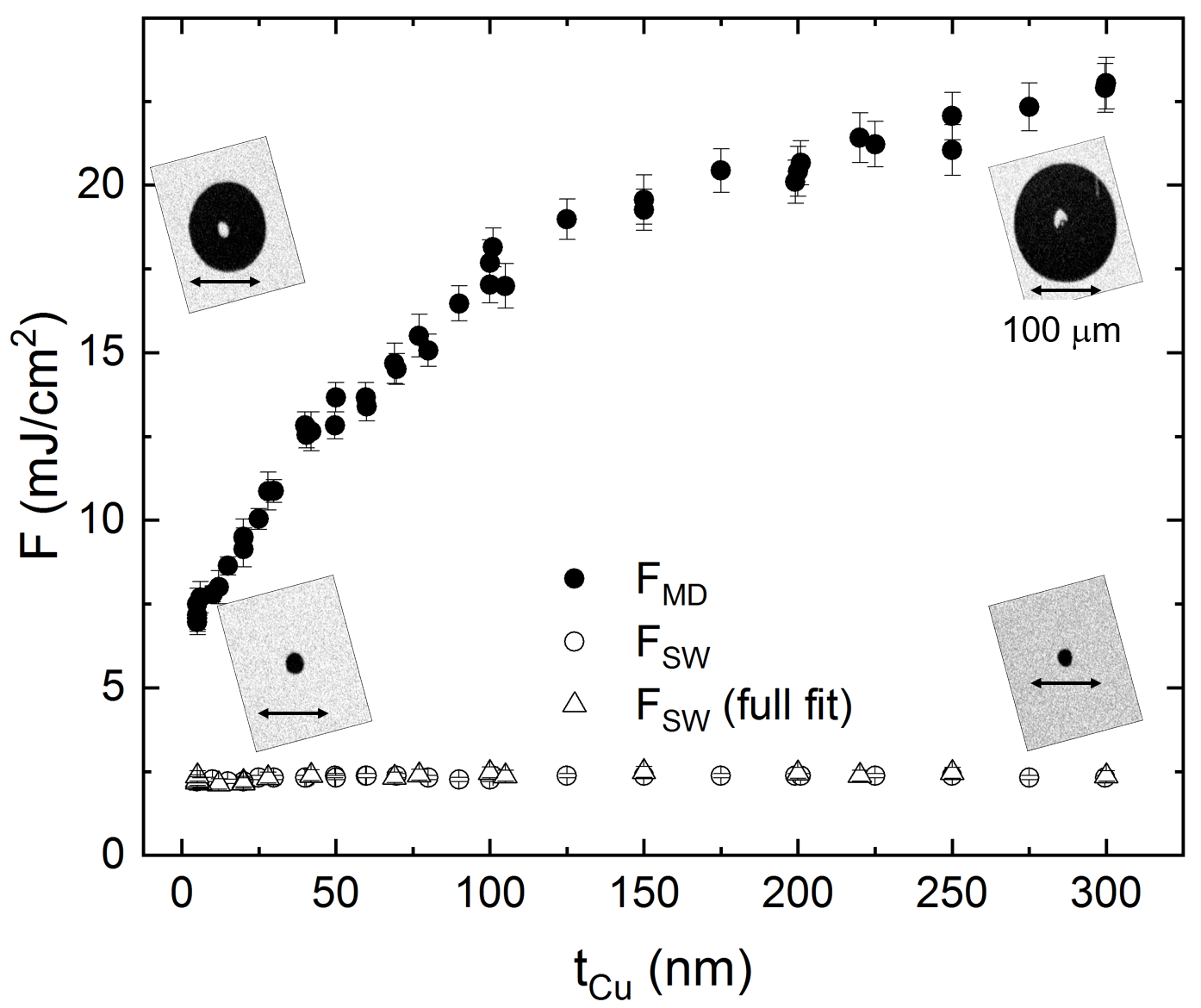}
\caption{\label{fig_Fvsd150fs} $F_{\rm SW}$ and $F_{\rm MD}$ vs the heat sink thickness, $t_{\rm Cu}$, for s-polarized, $750\,$nm laser pulses of $150\,$fs duration incident at $\approx 5\,$deg. Open triangles represent results of the fits used to determine the beam size. Insets show typical domains written at $F_{\rm SW}$ and $F_{\rm MD}$ for $t_{\rm Cu}=5$ and $300\,$nm, respectively.}
\end{figure}

The fluence thresholds $F_{\rm SW}$ and $F_{\rm MD}$ are derived from the size and quality of written domains imaged as function of $F_{\rm las}$, $\tau_{\rm las}$, and $t_{\rm Cu}$ by magneto-optical microscopy (cf.\,insets of Fig.\,\ref{fig_Fvsd150fs}). We first calibrate this derivation by simultaneously fitting the radii $r_{x/y}$ of the switched domains  observed in  samples with particular  heat sink thicknesses after their excitation by $p$-polarized, $800\,$nm, $150\,$fs laser pulses of various pulse energies, $E_{\rm las}$, to
\begin{equation}
r_{x/y}=\sigma_{x/y}\sqrt{\ln\left(\frac{E_{\rm las}}{\pi \sigma_x \sigma_y F_{\rm SW}}\right)} \,\,\, ,
\end{equation}
where $x$ and $y$ are orthogonal axis defining the sample plane and $\sigma_{x/y}$ denote the corresponding widths of our slightly elliptical, Gaussian laser beam on the sample. The laser spot area obtained from these fits is then used to convert the measured pulse energy thresholds for the appearance of reversed domains at other values of $t_{\rm Cu}$, and of non-reversed domains within the written ones into $F_{\rm SW}$ and $F_{\rm MD}$, respectively. 

The corresponding results,  shown in figure \ref{fig_Fvsd150fs} for $t_{\rm Cu} \in [5,300]\,$nm, reveal that $F_{\rm SW}$ is independent of $t_{\rm Cu}$, but that  $F_{\rm MD}$ strongly increases with $t_{\rm Cu}$ in a way that resembles the increase of femtosecond laser pulse damage thresholds of gold films with their thickness \cite{WHG_99}.  In analogy to the fact that the damage threshold of Au is governed by $T_{\rm e, eq}= T_{\rm melt}$, with $T_{\rm e,eq}$ and $T_{\rm melt}$ denoting the electron temperature at the time when they reach thermal equilibrium with the phonons and the melting point, respectively, this resemblance indicates that multi-domain formation is also governed by a threshold condition for $T_{\rm e}$ at longer time delays. 

\begin{figure}[ht] 
\hspace*{-.1cm}\includegraphics[width=8.9cm]{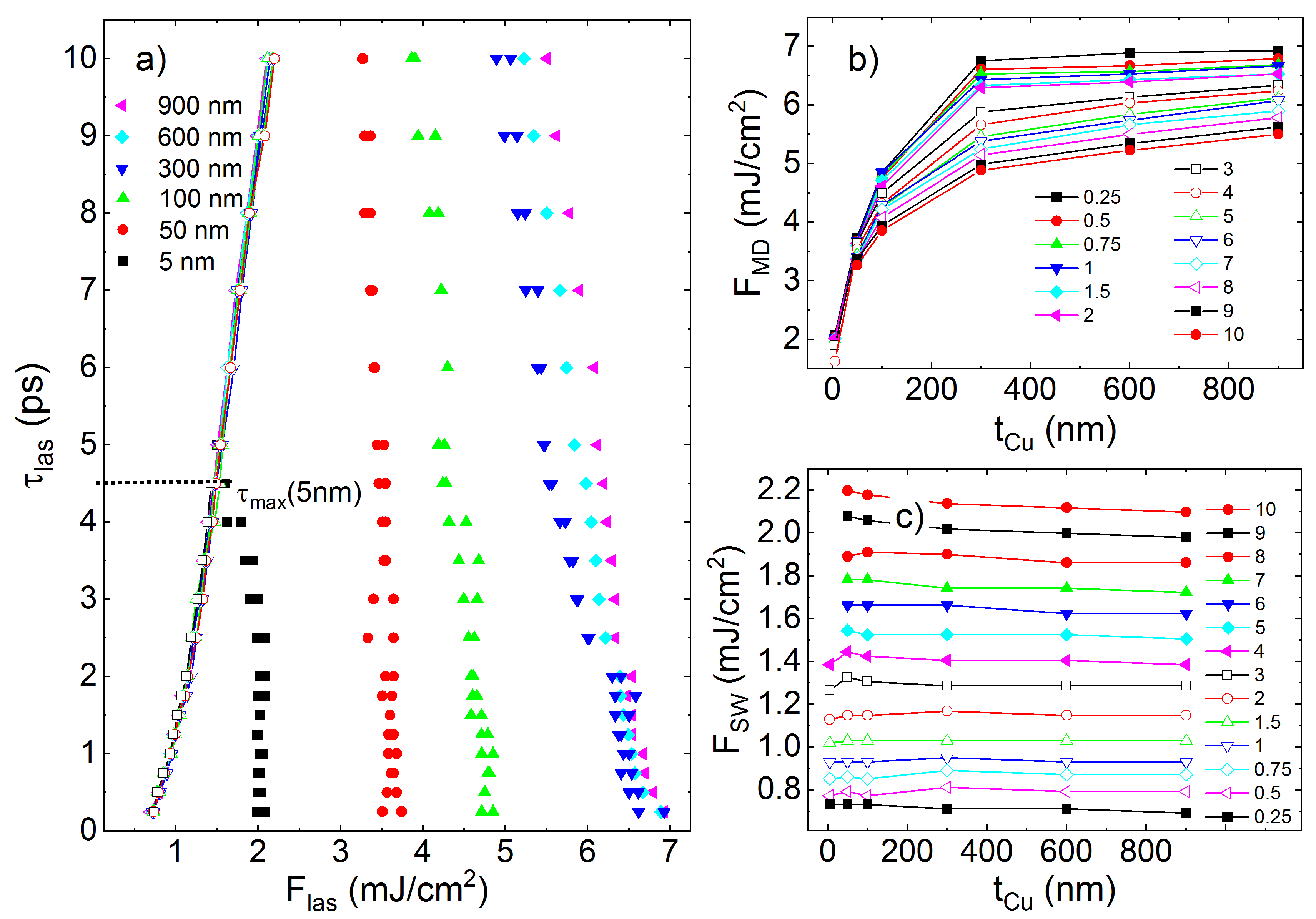}
\caption{\label{fig_statediagram} a) $F_{\rm SW}(\tau_{\rm las})$ (open symbols/lines) and $F_{\rm MD}(\tau_{\rm las})$ (closed symbols) for different $t_{\rm Cu}$ given in the legend. The dashed line indicates $\tau_{\rm max}(t_{\rm Cu}=5\,$nm). b) $F_{\rm MD}$ vs $t_{\rm Cu}$ for various $\tau_{\rm las}$ given in the legend in ps. c) Same as b) but for $F_{\rm SW}$. All data are for $p$-polarized, $800\,$nm pulses incident at $\approx 56\,$deg. }
\end{figure}

We repeated the procedure outlined above to determine the impact of the laser pulse duration on $F_{\rm SW}$ and $F_{\rm MD}$ for selected heat sink thicknesses shown in figure \ref{fig_statediagram}.
Throughout the investigated pulse duration range from $250\,$fs to $10\,$ps, we find that the values of $F_{\rm SW}$ monotonously increase with $\tau_{\rm las}$ but are independent of $t_{\rm Cu}$ (cf.\,panels a) and c)). A comparison of these findings to TTM calculations with typical material parameters as published in \cite{GWY_16, HWG_00,HDP_12} shows, that both results contradict the hypothesis of a $T_{\rm e,max}$-defined switching criterion for AO-HIS. In order to induce identical values of $T_{\rm e,max}$ in samples with $t_{\rm Cu}=5\,$nm, the fluence of $5\,$ps pulses would have to be about four times larger than the one of $250\,$fs pulses -- about twice as large as the measured ratio. Moreover, in order to compensate for cooling of the electrons due to heat diffusion into the heat sink during the excitation, $F_{\rm SW}$ would have to increase with $t_{\rm Cu}$ if the appearance of switching would be related to $T_{\rm e,max}$. According to the TTM, about $40\,$\% larger fluences are needed to reach the same $T_{\rm e,max}$ with $5\,$ps pulses in samples with $t_{\rm Cu}=200\,$nm as in those with $t_{\rm Cu}=5\,$nm. In line with the work of Gorchon {\it et al.}\,\cite{GWY_16}, the incompatibility of our data and TTM calculations strongly suggests  that $F_{\rm SW}$ is not directly related  to $T_{\rm e,max}$.

Now we draw our attention to the impact of $t_{\rm Cu}$ and $\tau_{\rm las}$ on $F_{\rm MD}$.  As shown in Fig.\,\ref{fig_statediagram} b), we find that $F_{\rm MD}$  increases with $t_{\rm Cu}$  in a way that resembles exponential saturation functions, $F_{\rm MD}=F_0+\Delta F \cdot  (1-\exp(-t_{\rm Cu}/\lambda_{\rm Cu}))$ with characteristic amplitudes, $\Delta F$, and lengths, $\lambda_{\rm Cu}$, that decrease and increase with $\tau_{\rm las}$, respectively. The behavior of $\Delta F$ is qualitatively reproduced by our TTM calculations and reflects the changing character of heat diffusion with increasing pulse duration.  While short pulses lead to thermal diffusion by hot electrons that rapidly spreads the absorbed energy over a range of $\lambda_{\rm th,e}\approx 230\,$nm before electrons and phonons reach thermal equilibrium, long pulses are followed by much slower phonon dominated diffusion that spreads the energy only over $\lambda_{\rm th,ph}\approx 60\,$nm within $10\,$ps \cite{WHG_99,CBS_88}. In contrast, the increase of $\lambda_{\rm Cu}$ with $\tau_{\rm las}$ (related to the slight uptick of $F_{\rm MD}$ with $t_{\rm Cu}\in [300,900]\,$nm observed for long pulse durations) cannot be explained in terms of temperature dynamics alone and might be caused by magnetic contributions to the formation of multiple domains \cite{OIM_09,ShK_86}.

The demonstration of a strong increase of the multi-domain threshold with the heat sink thickness constitutes the most important result of this work. It proves, independent of any model,  that values of $F_{\rm MD}$ reported for non-heat sunk samples do not reflect a breakdown of the material specific reversal mechanism due to an exceedingly high maximum electron temperature. Instead, the temperature of insufficiently cooled electrons becomes so large at $F_{\rm MD}$, that it leads to complete demagnetization and/or activates dipolar field driven multi-domain formation at times longer than the electron-phonon equilibration time \cite{OIM_09,ShK_86}. Moreover, it reveals that the maximum pulse duration applicable for controlled AO-HIS, $\tau_{\rm max}$, is not limited by accelerated demagnetization of Gd \cite{DJM_20} but by thermally activated multi-domain formation.  As obvious from figure \ref{fig_statediagram} a), $\tau_{\rm max}$ is defined by $F_{\rm SW}(\tau_{\rm max})=F_{\rm MD}(\tau_{\rm max})$, so that an increase of $F_{\rm MD}$ with $t_{\rm Cu}$  enlarges $\tau_{\rm max}$. The impact of $F_{\rm MD}(t_{\rm Cu})$ on $\tau_{\rm max}$ is so strong, that $\tau_{\rm max}$ already exceeds our experimental pulse length limit of $10\,$ps for $t_{\rm Cu}=50\,$nm. Linear data extrapolation suggests, that pulses as long as $25\,$ps could be used for AO-HIS in samples with $t_{\rm Cu}=900\,$nm.  

So far, we used static images of magnetization patterns resulting from the illumination of {\it multiple samples} (with distinct $t_{\rm Cu}$) by laser pulses of {\it various fluences} to prove that $F_{\rm MD}$ is not governed by a breakdown of the reversal mechanism of AO-HIS. Now, we show that investigations of the spatiotemporal magnetization changes within a {\it single sample} induced by laser pulses of {\it one fixed fluence} might prove this fact as well. To provide such proof, the time-resolved images need to demonstrate that laser pulse excitation with $F_{\rm las}\ge F_{\rm MD}$ are followed by spatially continuous magnetization reversal before the magnetization breaks up into multiple domains. Consequently, the time-resolved approach works only for samples where large dipolar fields and/or small magneto-crystalline anisotropies lead to the formation multi-domains at temperatures so far below the Curie point, that the laser heated electrons at no time  cause complete demagnetization. This implies that the multi-domain thresholds for femtosecond laser pulse excitation of suitable samples do not exceed $F_{\rm SW}$ by much, if at all. Since this criterion is not fulfilled for the samples investigated so far (they exhibit $F_{\rm MD}/F_{\rm SW} \ge 3$), we demonstrate the viability of the time-resolved approach by example of a  glass/Ta(5)/Pt(3)/$\alpha$-Gd$_{18}$Dy$_4$Co$_{78}$(10)/Pt(5) sample whose multi-domain threshold is just $\approx 29\,$\% larger than $F_{\rm SW}=4.6\,$mJ/cm$^2$ for $s$-polarized $150\,$fs, $800\,$nm pulses incident at $\approx 5\,$deg.

\begin{figure}[ht] 
\hspace*{-.1cm}\includegraphics[width=8.7cm]{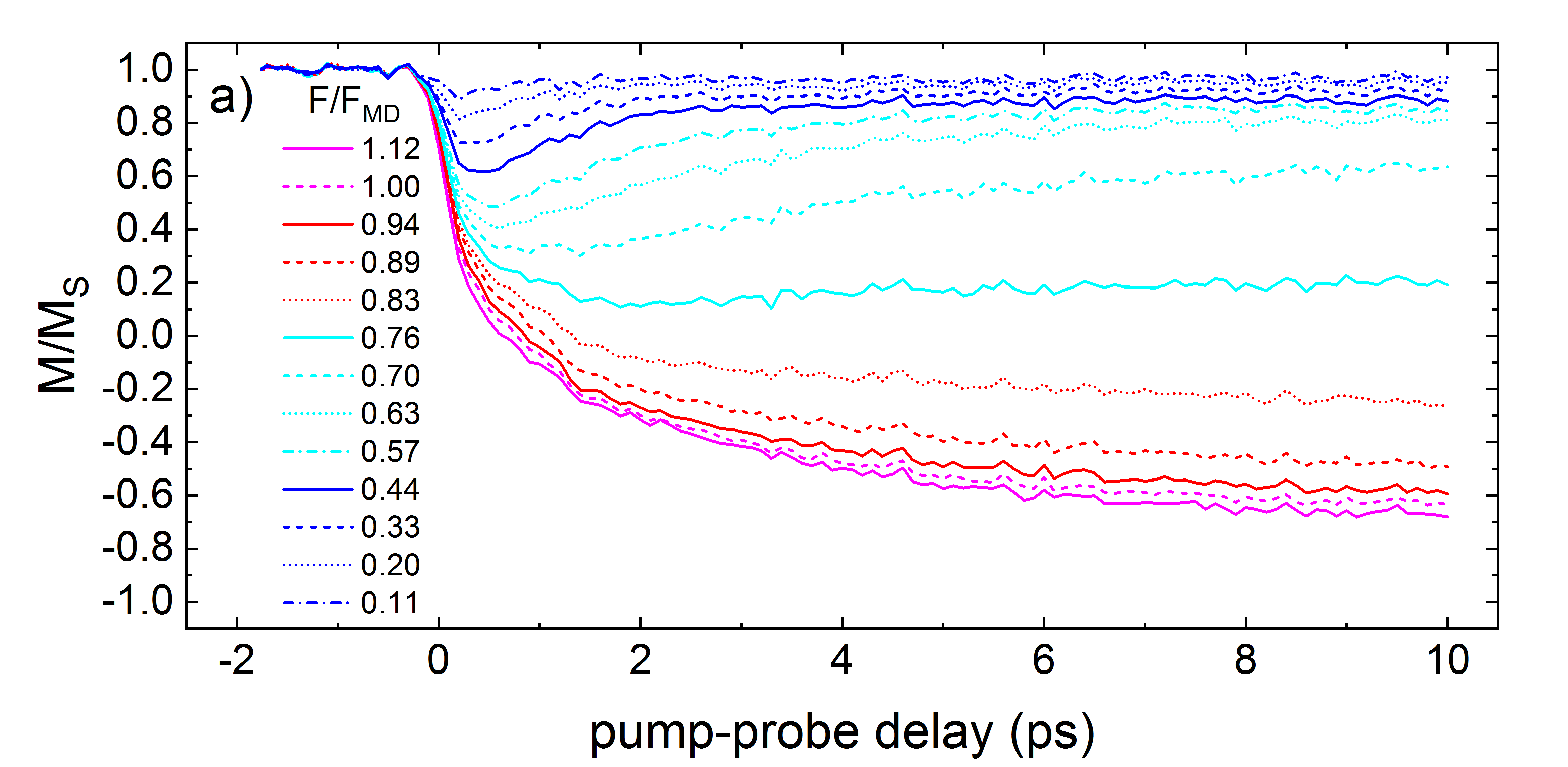}
\hspace*{.0cm}\includegraphics[width=8.7cm]{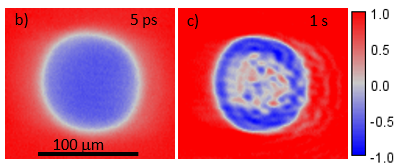}
\caption{\label{fig_magdyn} a) Magnetization dynamics in $\alpha$-Gd$_{18}$Dy$_4$Co$_{78}$ induced by s-polarized, $150\,$fs, $800\,$nm laser pulses of various fluences (given in multiples of $F_{\rm MD}=5.9\,$mJ/cm$^2$) incident at $\approx 5\,$deg. Panels b) and c) show normalized magnetization profiles $5\,$ps and $1\,$s after the excitation, respectively.}
\end{figure}

 Both, the fluence dependent magnetization dynamics shown in figure\,\ref{fig_magdyn}a), and the magnetization patterns measured at short and long delays given in figures\,\ref{fig_magdyn}b) and c), respectively, clearly attest that the excitation of this sample by $150\,$fs laser pulses with $F_{\rm las} = 1.12 \cdot F_{\rm MD}$ is followed by ultrafast, continuous magnetization reversal before elevated electron temperatures in quasi-equilibrium with the phonon temperature activate the dipolar field driven formation of multiple domains.   

In conclusion we have used the thickness of Cu heat sinks to control the electron temperature dynamics following optical excitation of adjacent GdFeCo films by laser pulses of various duration and fluence. This enabled us to separately study the impact of the electron temperature at short times, where it reaches its maximum, and at longer times, where it is in quasi-equilibrium with the phonon temperature, on the fluence thresholds for the onset of controlled temperature induced switching and of stochastic multi-domain formation. The measured variations of these thresholds with the heat sink thickness and laser pulse duration evidence that neither one is solely governed by $T_{\rm e,max}$. We have shown that $F_{\rm MD}$ corresponds to the fluence where the temperature of insufficiently cooled electrons  becomes so large that it causes complete demagnetization and/or activates dipolar field driven multi-domain formation at times  longer than the electron-phonon equilibration time. Based on this new insight we predict that excitation of samples whose multi-domain threshold does not exceed the switching threshold by much, if at all, with laser fluences slightly larger than $F_{\rm MD}$ lead to controlled switching before the magnetization breaks up into multiple domains. We confirmed this prediction by example of Gd$_{18}$Dy$_4$Co$_{78}$, with $F_{\rm MD}/F_{\rm SW}\approx 1.29$ and $F_{\rm las}=1.12 \cdot F_{\rm MD}$. Last, but not least, we have shown that any increase of $F_{\rm MD}$ with the heat sink thickness lengthens  the maximum pulse duration applicable for AO-HIS and derived an upper limit of $\tau_{\rm max }\approx 25\,$ps for Gd$_{24}$(Fe$_{90}$Co$_{10})_{76}$ from linear extrapolation of our data. We expect that thermal management by metallic heat sinks will not only pave the way for future technological applications of AO-HIS with laser pulses emitted by cheap laser diodes, but will also lead to the demonstration of AO-HIS in many ferrimagnetic alloys and multi-layers that were believed to show only multi-domain formation as already demonstrated for Gd$_{13}$Dy$_7$Co$_{80}$ in \cite{ZHH_23}.

\end{document}